\pdfoutput=1

\documentclass[runningheads]{llncs}
\usepackage{graphicx}
\usepackage{acro}
\usepackage{amsmath}
\usepackage{amssymb}
\usepackage{hyperref}

\DeclareAcronym{US}{
short=US,
long=ultrasound
}

\DeclareAcronym{HVF}{
short=GHIF,
long=global hetero-image fusion
}

\DeclareAcronym{NN}{
short=CNN,
long=convolutional neural network
}

\DeclareAcronym{VSP}{
short=VSP,
long=view-specific parameterization
}

\DeclareAcronym{HeMIS}{
short=HeMIS,
long=hetero-modal image segmentation
}

\DeclareAcronym{ROI}{
short=ROI,
long=region of interest
}

\DeclareAcronym{FCN}{
short=FCN,
long=fully-convolutional network
}

\DeclareAcronym{AUC}{
short=AUC,
long=area under the curve,
long-plural-form=areas under the curve,
}

\DeclareAcronym{ROC}{
short=ROC,
long=receiver operating characteristic,
}

\DeclareAcronym{R@P90}{
short=R@P90,
long=recall at 90\% precision,
}
\DeclareAcronym{R@P80}{
short=R@P80,
long=recall at 80\% precision,
}
\DeclareAcronym{R@P85}{
short=R@P85,
long=recall at 85\% precision,
}
\newcommand{\etal}{\textit{et al.}}
\newcommand{\eg}{\textit{e.g.},}
\newcommand{\ie}{\textit{i.e.},}

\begin{document}
\title{Reliable Liver Fibrosis Assessment from Ultrasound using Global Hetero-Image Fusion and View-Specific Parameterization}
\titlerunning{Liver Fibrosis Assessment from Ultrasound using GHIF and VSP}
% \author{Paper Submission 823\#}

\author{Bowen Li\inst{1} \and Ke Yan\inst{1} \and Dar-In Tai\inst{2} \and Yuankai Huo\inst{3} \and Le Lu\inst{1} \and Jing Xiao\inst{4} \and Adam P. Harrison\inst{1}}
% Index{Li, Bowen}
% Index{Yan, Ke}
% Index{Tai, Dai-In}
% Index{Huo, Yuankai}
% Index{Lu, Le}
% Index{Xiao, Jing}
% Index{Harrison, Adam}

\authorrunning{B.Li et al.}

\institute{PAII Inc., Bethesda, MD 20817, USA \and
Chang Gung Memorial Hospital, Linkou, Taiwan, ROC \and
Vanderbilt University, Nashville, TN 37235, USA \and
PingAn Technology, Shenzhen, China}
\maketitle              % typeset the header of the contribution
\begin{abstract}
\Ac{US} is a critical modality for diagnosing liver fibrosis. Unfortunately, assessment is very subjective, motivating automated approaches. We introduce a principled deep \ac{NN} workflow that incorporates several innovations. First, to avoid overfitting on non-relevant image features, we force the network to focus on a clinical \ac{ROI}, encompassing the liver parenchyma and upper border. Second, we introduce \ac{HVF}, which allows the \ac{NN} to fuse features from any arbitrary number of images in a study, increasing its versatility and flexibility. Finally, we use ``style''-based \ac{VSP} to tailor the \ac{NN} processing for different viewpoints of the liver, while keeping the majority of parameters the same across views. Experiments on a dataset of $610$ patient studies ($6979$ images) demonstrate that our pipeline can contribute roughly $7\%$ and $22\%$ improvements in partial \acl{AUC} and \acl{R@P90}, respectively, over conventional classifiers, validating our approach to this crucial problem. 

\end{abstract}
\acresetall
\section{Introduction}

Liver fibrosis is a major health threat with high prevalence~\cite{poynard2010prevalence}. Without timely diagnosis and treatment, liver fibrosis can develop into liver cirrhosis~\cite{poynard2010prevalence} and even hepatocellular carcinoma~\cite{saverymuttu1986ultrasound}. While histopathology remains the gold standard, non-invasive approaches minimize patient discomfort and danger. Elastography is a useful non-invasive modality, but it is not always available or affordable and it can be confounded by inflammation, presence of steatosis, and the patient's etiology~\cite{Tai_2015,Chen_2019,Lee_2017}. Assessment using conventional \ac{US} may be potentially more versatile; however, it is a subjective measurement that can suffer from insufficient sensitivities, specificities, and high inter- and intra-rater variability~\cite{Manning_2008,Li_2019}. Thus, there is great impetus for an automated and less subjective assessment of liver fibrosis. This is the goal of our work. 

%Conventional ultrasound is a major modality being used to diagnose liver fibrosis[?], which is inexpensive, non-invasive and can be interpreted in real time. However, reading liver ultrasound images requires highly experienced readers, and the diagnosis criteria are subjective, which results in a low inter-reader reliability[?]. Thus, a clinical decision support system is needed for the liver ultrasound reading scenario.

Although a relatively understudied topic, prior work has advanced automated \ac{US} fibrosis assessment~\cite{Mojsilovic_1997,Wu_1992,Ogawa_1998,meng2017liver,liu2019ultrasound}. In terms of deep \acp{NN}, Meng \etal{}~\cite{meng2017liver} proposed a straightforward liver fibrosis parenchyma VGG-16-based~\cite{Simonyan15} classifier and tested it on a small dataset of $279$ images. Importantly, they only performed image-wise predictions and do not report a method for study-wise classification.  On the other hand, Liu \etal{}~\cite{liu2019ultrasound} correctly identified the value in fusing features from all \ac{US} images in a study when making a prediction. However, their algorithm requires exactly $10$ images. But, real patient studies may contain any arbitrary number of \ac{US} scans. Their feature concatenation approach would also drastically increase computational and memory costs as more images are incorporated. Moreover, they rely on $13$ manually labeled indicators as ancillary supervision, which are typically not available without considerable labor costs. Finally, their system treats all \ac{US} images identically, even though a study consists of different \textit{viewpoints} of the liver, each of which may have its own set of clinical markers correlating with fibrosis. Ideally, a liver fibrosis assessment system could learn directly from supervisory signals already present in hospital archives, \ie{} image-level fibrosis scores produced during daily clinical routines. In addition, a versatile system should also be able to effectively use all \ac{US} images/views in a patient study with no ballooning of computational costs, regardless of their number. 

We fill these gaps by proposing a robust and versatile pipeline for conventional ultrasound liver fibrosis assessment. Like others, we use a deep \ac{NN}, but with key innovations. First, informed by clinical practice~\cite{Aube_2017}, we ensure the network focuses only on a clinically-important \ac{ROI}, \ie{} the liver parenchyma and the upper liver border. This prevents the \ac{NN} from erroneously overfitting on spurious or background features. Second, inspired by \ac{HeMIS}~\cite{Havaei_2016}, we adapt and expand on this approach and propose \ac{HVF} as a way to learn from, and perform inference on, any arbitrary number of \ac{US} scans within a study. While \ac{HVF} share similarities with deep feature-based multi-instance learning~\cite{ilse2018attention}, there are two important distinctions: (1) \ac{HVF} includes variance as part of the fusion, as per \ac{HeMIS}~\cite{Havaei_2016}; (2) \ac{HVF} is trained using arbitrary image combinations from a patient study, which is possible because, unlike multi-instance learning, each image (or instance) is strongly supervised by the same label. We are the first to propose and develop this mechanism to fuse global \ac{NN} feature vectors.  Finally, we implement a \ac{VSP} that tailors the \ac{NN} processing based on $6$ common liver \ac{US} views. While the majority of processing is shared, each view possesses its own set of so-called ``style''-based normalization parameters~\cite{Huang_2017} to customize the analysis. While others have used similar ideas segmenting different anatomical structures~\cite{Huang_2019}, we are the first to apply this concept for clinical decision support and the first to use it in concert with a hetero-image fusion mechanism. The result is a highly robust and practical liver fibrosis assessment solution.

To validate our approach, we use a cross-validated dataset of $610$ US patient studies, comprising $6976$ images. We measure the ability to identify patients with moderate to severe liver fibrosis. Compared to strong classification baselines, our enhancements are able to improve recall at $90\%$ precision by $22\%$, with commensurate boosts in partial \acp{AUC}.  Importantly, ablation studies demonstrate that each component contributes to these performance improvements, demonstrating that our liver fibrosis assessment pipeline, and its constituent clinical \ac{ROI}, \ac{HVF}, and \ac{VSP} parts, represents a significant advancement for this important task. 

 %Furthermore, we demonstrate that the performance of our approach degrades gracefully as the number of \ac{US} scans given to it are reduced during inference. This is in contrast to baseline approaches.

%\{A paragraph about view fusion, place holder\}

%To overcome all the obstacles mentioned, and to help clinicians make more objective, information-empowered diagnosis, we put forward an end-to-end solution for liver fibrosis prediction with conventional ultrasound images. It does not require a fixed amount of images, and does not require additional training labels, instead, it can take a random set of ultrasound images as input, and predict whether the subject liver has fibrosis.

%\{Contributions: A view fusion method; A model for liver fibrosis prediction., place holder\}

\section{Methods}

\begin{figure}[t]
\includegraphics[width=\textwidth]{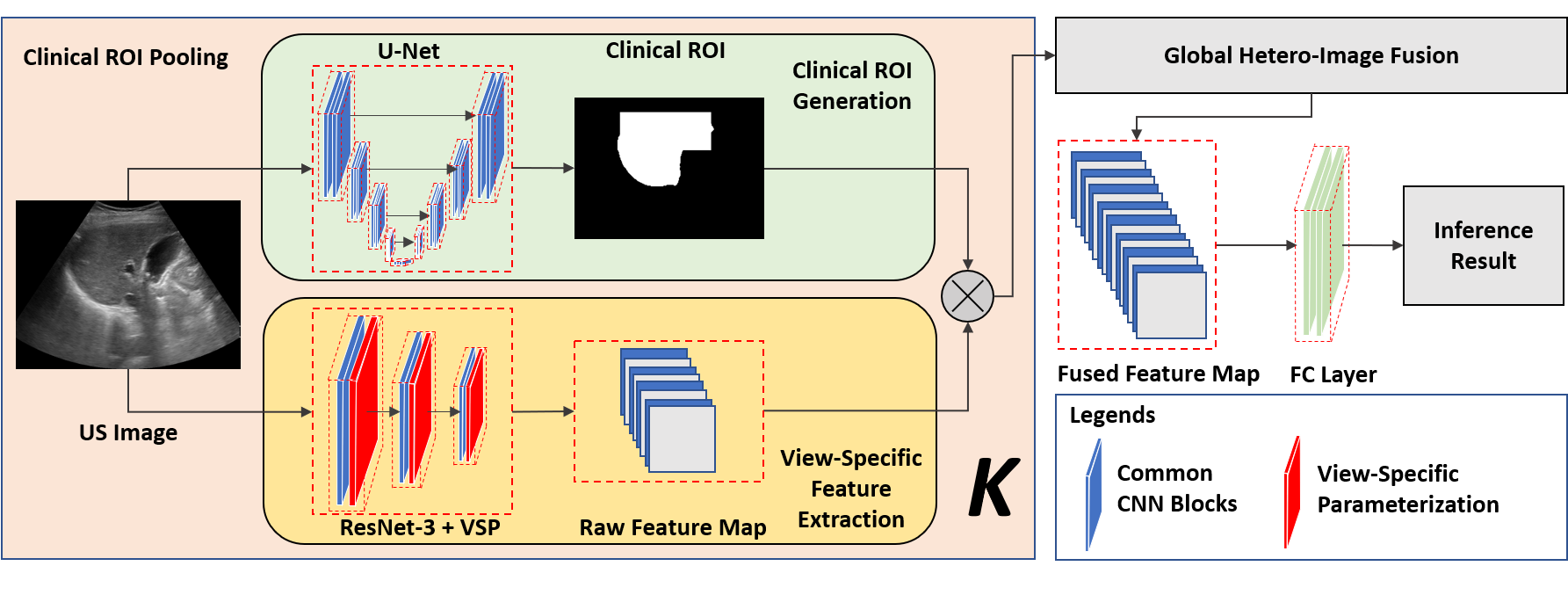}
\caption{Algorithmic workflow depicting the clinical \acs{ROI} pooling, \acs{HVF}, and \acs{VSP}. We use plate notation to depict the repeated workflow across the $K$ images in a \ac{US} study.} \label{fig-alg} 
\end{figure}

We assume we are given a dataset, $\mathcal{D}=\{\mathcal{X}_{i}, y_{i}\}_{i=1}^{N}$, comprised of \ac{US} patient studies and ground-truth labels indicating liver fibrosis status, dropping the $i$ when convenient. Each study $\mathcal{X}_{i}$, in turn, is comprised of an arbitrary number of $K_{i}$ 2D conventional \ac{US} scans of the patient's liver, $\mathcal{X}_{i}=\{\mathbf{X}^{1}\ldots \mathbf{X}^{K_{i}}\}$. Fig.~\ref{fig-alg} depicts the workflow of our automated liver assessment tool, which combines clinical \ac{ROI} pooling, \ac{HVF}, and \ac{VSP}. 

\subsection{Clinical ROI Pooling}

We use a deep \ac{NN} as backbone for our pipeline. Popular deep \acp{NN}, \eg{} ResNet~\cite{He_2016}, can be formulated with the following convention:
\begin{align}
    \hat{y}^{k}  &= f\left(g\left(\mathbf{A}^{k} \right); \mathbf{w}\right) \mathrm{,} \label{eqn:simple_image} \\
    \mathbf{A}^{k}&=h\left(\mathbf{X}^{k}; \theta\right) \mathrm{,} \label{eqn:simple_image2}
\end{align}
where $h(.; \theta)$ is a \ac{FCN} feature extractor parameterized by $\theta$, $\mathbf{A}^{k}$ is the \ac{FCN} output, $g(.)$ is some global pooling function, \eg{} average pooling, and $f(.; \mathbf{w})$ is a fully-connected layer (and sigmoid function) parameterized by $\mathbf{w}$. When multiple \ac{US} scans are present, a standard approach is to aggregate individual image-wise predictions, \eg{} taking the median:
\begin{align}
    \hat{y} = \mathrm{median}(\{\hat{y}^{1}\ldots \hat{y}^{K}\}) \mathrm{.} \label{eqn:median}
\end{align}

\begin{figure}[t]
\center
\begin{tabular}{ccc}
     \includegraphics[width=.25\textwidth]{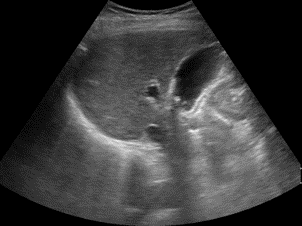} & \includegraphics[width=.25\textwidth]{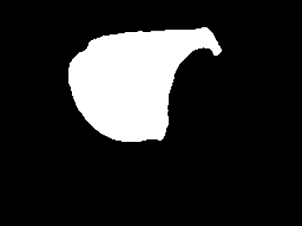} & \includegraphics[width=.25\textwidth]{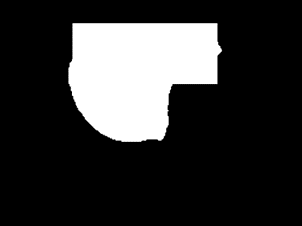} \\
     (a) \textbf{US Image} & (b) \textbf{Liver Mask} & (c) \textbf{Clinical ROI}
\end{tabular}
\caption{(a) depicts an \ac{US} image, whose liver mask is rendered in (b). As shown in (c), the clinical \ac{ROI} the mask is extended upward to cover the top liver border.}
\label{fig:roi}
\end{figure}

This conventional approach may have drawbacks, as it is possible for the \ac{NN} to overfit to spurious background variations. However, based on clinical practice~\cite{Aube_2017}, we know \textit{a priori} that certain features are crucial for assessing liver fibrosis, \eg{} the parenchyma texture and surface nodularity. As Fig.~\ref{fig:roi} demonstrates, to make the \ac{NN} focus on these features we use a masking technique.  We first generate a liver mask for each \ac{US} scan. This is done by training a simple segmentation network on a small subset of the images.
Then, for each scan, we create a rectangle that just covers the top half of and 10 pixels above the liver mask, to ensure the liver border is covered. The resulting binary mask is denoted $\mathbf{M}$. Because we only need to ensure we capture enough of the liver parenchyma and upper border to extract meaningful features, $\mathbf{M}$ need not be perfect. 

With a clinical \ac{ROI} obtained, we formulate the pooling function in \eqref{eqn:simple_image} as a masked version of global average pooling:
\begin{align}
    g(\mathbf{A}; \mathbf{M}) = \mathrm{GAP}(\mathbf{M}\odot\mathbf{A}) \mathrm{,} \label{eqn:roi_pool}
\end{align}
where $\odot$ and $\mathrm{GAP}(.)$ denote the element-wise product and global average pooling, respectively. Interestingly, we found that including the zeroed-out regions within the global average pooling benefits performance~\cite{chen2020anatomyaware,pelvic_yirui}. We posit their inclusion helps implicitly capture liver size characteristics, which is another important clinical \ac{US} marker for liver fibrosis~\cite{Aube_2017}. 

%\sigma\left(\mathbf{w}^{\intercal}

%For each selected $image_i$ of view group $v_i$, first it goes through a pre-trained unet to get the segmentation of the liver $seg_i$. A square area $s_i$ is calculated to cover the top half of $seg_i$, and the top boundary of $s_i$ is moved upwards by 10 pixels. The region of interest $ROI_i$ is calculated as: $$ROI_i = seg_i \cup s_i$$

%A raw feature map $RFM_i$ is also calculated for $image_i$, with a modified resnet. The basis of the modified resnet is resnet50, but with the following modification: a) only the first three layers are used; b) all batch normalization layers are substituted by a set of view-specific trainable instance normalization layers. The modified reset is called $Resnet-3i-v_i$.

%To get a pooled feature map $PFM_i$, a ROI pooling is then applied to $RFM_i$: $$PFM_i=RFM_i\times ROI_i$$

\subsection{Global Hetero-Image Fusion}

A challenge with \ac{US} patient studies is that they may consist of a variable number of images, each of a potentially different view. Ideally, all available \ac{US} images would contribute to the final prediction. In \eqref{eqn:median} this is accomplished via a late fusion of independent and image-specific predictions. But, this does not allow the \ac{NN} integrate the combined features across \ac{US} images. A better approach would fuse these features. The challenge, here, is to allow for an arbitrary number of \ac{US} images in order to ensure flexibility and practicality. 

The \ac{HeMIS} approach~\cite{Havaei_2016} to segmentation offers a promising strategy that fuses features from arbitrary numbers of images using their first- and second-order moments. However, \ac{HeMIS} fuses convolutional features early in its \ac{FCN} pipeline, which is possible because it assumes pixel-to-pixel correspondence across images. This is completely violated for \ac{US} images. Instead, only \textit{global} \ac{US} features can be sensibly fused together, which we accomplish through \acf{HVF}. More formally, we use $\mathcal{A}= \{\mathbf{A}^{k}\}_{k=1}^{K}$ and $\mathcal{M}= \{\mathbf{M}^{k}\}_{k=1}^{K}$ to denote the set of \ac{FCN} features and clinical \acp{ROI}, respectively, for each image. Then \ac{HVF} modifies \eqref{eqn:simple_image} to accept any arbitrary set of \ac{FCN} features to produce a \textit{study-wise} prediction:
\begin{align}
    \hat{y}  &= f\left(g\left( \mathcal{A}; \mathcal{M}\right); \mathbf{w}\right) \mathrm{,} \label{eqn:all_hvf} \\
    g(\mathcal{A}; \mathcal{M}) &= \mathrm{concat}\left(\mathrm{mean}(\mathcal{G}), \mathrm{var}(\mathcal{G}), \max(\mathcal{G})\right) \mathrm{,} \label{eqn:hvf}\\
    \mathcal{G} &= \{ \mathrm{GAP}(\mathbf{M}^{k}\odot\mathbf{A}^{k})\}_{k=1}^{K} \mathrm{,}
\end{align}
Besides the first- and second-order moments, \ac{HVF} \eqref{eqn:hvf} also incorporates the max operator as a powerful hetero-fusion function~\cite{Zhou_2018_fusion}. All three operators can accept any arbitrary numbers of samples to produce one fused feature vector. To the best of our knowledge, we are the first to apply hetero-fusion for global feature vectors. The difference, compared to late fusion, is that features, rather than predictions are fused. Rather than always inputting all \ac{US} scans when training, an important strategy is choosing random combinations of the $K$ scans for every epoch. This provides a form of data augmentation and allows the \ac{NN} to learn from image signals that may be suppressed otherwise. An important implementation note is that training with random combinations of images can make \ac{HVF}'s batch statistics unstable. For this reason, a normalization not relying on batch statistics, such as instance-normalization~\cite{Ulyanov_2016}, should be used.  

% Here, the \ac{HVF} of \eqref{eqn:hvf} also incorporates the max operator, in addition to the first- and second-order moments, as that was shown to also be a powerful hetero-fusion function~\cite{Zhou_2018_fusion}, which we found can enhance performance. 

\subsection{View-Specific Parameterization}

\begin{figure}[t]
\centering
    \includegraphics[width=1.0\textwidth]{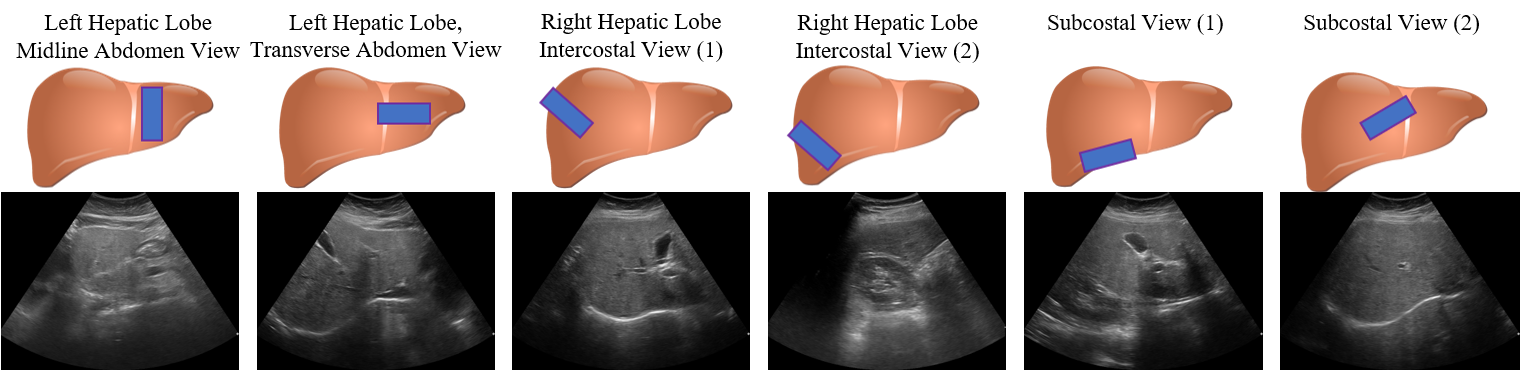}
    \caption{Liver views in our dataset. Blue square: position of the \ac{US} probe. Liver cartoons adapted from the DataBase Center for Life Science (\url{https://commons.wikimedia.org/wiki/File:201405_liver.png}), licensed under the  Creative Commons Attribution 4.0 International}
    \label{fig-view}
\end{figure}

While \ac{HVF} can effectively integrate arbitrary numbers of \ac{US} images within a study, it uses the same \ac{FCN} feature extractor across all images, treating them all identically. Yet, there are certain \ac{US} features, such as vascular markers, that are specific to particular views. As a result, some manner of view-specific analysis could help push performance further. In fact, based on guidance from our clinical partner, \ac{US} views of the liver can be roughly divided into $6$ categories, which focus on different regions of the liver. These are shown in Fig~\ref{fig-view}. 

%\footnotetext{Adapted from the DataBase Center for Life Science's image (\url{https://commons.wikimedia.org/wiki/File:201405_liver.png}), licensed under the  Creative Commons Attribution 4.0 International (\url{https://creativecommons.org/licenses/by/4.0/deed.en})}
%\footnote{Adapted from the DataBase Center for Life Science's image (\url{https://commons.wikimedia.org/wiki/File:201405_liver.png}), licensed under the  Creative Commons Attribution 4.0 International (\url{https://creativecommons.org/licenses/by/4.0/deed.en})}
% \begin{figure}[t]
% \center
% \begin{tabular}{cccccc}
%      \includegraphics[width=\cwidth]{figures/new_fig_3/1.png} & \includegraphics[width=\cwidth]{figures/new_fig_3/2.png} &  
%      \includegraphics[width=\cwidth]{figures/new_fig_3/3.png} & \includegraphics[width=\cwidth]{figures/new_fig_3/4.png} &
%      \includegraphics[width=\cwidth]{figures/new_fig_3/5.png} & \includegraphics[width=\cwidth]{figures/new_fig_3/6.png} \\
%       \includegraphics[width=\cwidth]{figures/1_2_img.png} & \includegraphics[width=\cwidth]{figures/3_4_img.png} &  
%      \includegraphics[width=\cwidth]{figures/5_6_img.png} & \includegraphics[width=\cwidth]{figures/9_10_img.png} &
%      \includegraphics[width=\cwidth]{figures/11_12_img.png} & \includegraphics[width=\cwidth]{figures/13_14_img.png}
% \end{tabular}
% %\includegraphics[width=\textwidth]{figures/figure-viewgroup.png}
% \caption{Liver views in our dataset. From left to right: left hepatic lobe, midline abdomen view; left hepatic lobe, transverse abdomen view; right hepatic lobe, intercostal view 1; right hepatic lobe, intercostal view 2; subcostal view 1; and subcostal view 2.} \label{fig-view}
% \end{figure}

A naive solution would be to use a dedicated deep \ac{NN} for each view category. However, this would drastically reduce the training set for each dedicated \ac{NN} and would sextuple the number of parameters, computation, and memory consumption. Intuitively, there should be a great deal of analysis that is common across \ac{US} views. The challenge is to retain this shared analysis, while also providing some tailored processing for each category. 

To do this, we adapt the concept of ``style'' parameters to implement a \acf{VSP} appropriate for \ac{US}-based fibrosis assessment. Such parameters refer to the affine normalization parameters used in batch-~\cite{Ioffe_2015} or instance-normalization~\cite{Ulyanov_2016}. If these are switched out, while keeping all other parameters constant, one can alter the behavior of the \ac{NN} in quite dramatic ways~\cite{Huang_2017,Huang_2019}. For our purposes, retaining view-specific normalization parameters allows for the majority of parameters and processing to be shared across views. \ac{VSP} is then realized with a minimal number of additional parameters. 

More formally, if we create $6$ sets of normalization parameters for an \ac{FCN}, we can denote them  as $\Omega=\{\omega_{1} \ldots \omega_{6}\}$. The \ac{FCN} from \eqref{eqn:simple_image2} is then modified to be parameterized also by $\Omega$:
\begin{align}
    \mathbf{A}^{k}&=h\left(\mathbf{X}^{k}; \theta, \omega_{v_{k}}\right) \mathrm{,} \label{eqn:vsp}
\end{align}
where $v_{k}$ indexes each image by its view and $\theta$ now excludes the normalization parameters. \ac{VSP} relies on identifying the view of each \ac{US} scan in order to swap in the correct normalization parameters. This can be recorded  as part of the acquisition process. Or, if this is not possible, we have found classifying the \ac{US} views automatically to be quite reliable.

\section{Experiments}

\textbf{Dataset.} We test our system on a dataset of $610$ \ac{US} patient studies collected from the Chang Gung Memorial Hospital in Taiwan, acquired from Siemens, Philips, and Toshiba devices. The dataset comprises $232$ patients, among which $95$ ($40.95\%$) patients have moderate to severe fibrosis (27 with severe liver steatosis). All patients were diagnosed with hepatitis B. Patients were scanned up to $3$ times, using a different scanner type each time. Each patient study is composed of up to $14$  \ac{US} images, corresponding to the views in Fig.~\ref{fig-view}. The total number of images is  $6\,979$. We use $5$-fold cross validation, splitting each fold at the patient level into $70\%$, $20\%$, and $10\%$, for training, testing, and validation, respectively. We also manually labeled liver contours from $300$ randomly chosen \ac{US} images.

\textbf{Implementation Details and Comparisons.} Experiments evaluated our workflow against several strong classification baselines, where throughout we use the same ResNet50~\cite{He_2016} backbone (pretrained on ImageNet~\cite{imagenet_cvpr09}). For methods using the clinical \ac{ROI} pooling of \eqref{eqn:roi_pool}, we use a truncated version of ResNet (only the first three layer blocks) for $h(.)$ in \eqref{eqn:simple_image2}. This keeps enough spatial resolution prior to the masking in \eqref{eqn:roi_pool}. We call this truncated backbone ``ResNet-3''. To create the clinical \ac{ROI}, we train a simple 2D U-Net~\cite{Ronneberger_2015} on the $300$ images with masks. For training, we perform standard data augmenation with random brightness, contrast, rotations, and scale adjustments. We use the stochastic gradient descent optimizer and a learning rate of 0.001 to train all networks. 
\begin{table}[t]
\setlength{\tabcolsep}{5pt}
\centering
\caption{Ablation Studies}\label{tab1e-ablation}
\begin{tabular}{cccccc}
\hline
\textbf{Method} &  \textbf{Partial AUC}&  \textbf{AUC} & \textbf{R@P90} & \textbf{R@P85} & \textbf{R@P80}\\
\hline
\hline
ResNet50 & $0.710$ & $0.893$ & $41.0\%$  & $61.7\%$  & $74.6\%$ \\
\hline
Clinical ROI & $0.744$ & $0.908$ & $46.1\%$ & $59.2\%$  & $82.1\%$ \\
\hline
Global Fusion & $0.591$ & $0.845$ & $30.2\%$ & $36.0\%$  & $49.8\%$\\
\hline
GHIF & $0.691$ & $0.885$ & $36.7\%$ & $41.0\%$  & $65.0\%$\\
\hline
GHIF (I-Norm) & $0.762$ & $0.907$ & $57.1\%$ & $71.1\%$  & $80.6\%$\\
\hline
GHIF + VSP (I-Norm) & $\mathbf{0.783}$ & $\mathbf{0.913}$ & $\mathbf{63.4\%}$  & $\mathbf{78.3\%}$  & $\mathbf{84.2\%}$\\
\hline
\end{tabular}
\end{table}

For baselines that can output only image-wise predictions, we test against a conventional ResNet50 and also a ResNet-3 with clinical \ac{ROI} pooling. For these two approaches, following clinical practices, we take the median value across the image-wise predictions to produce a study-wise prediction. All subsequent study-wise baselines are then built off the ResNet-3 with clinical \ac{ROI} pooling. We first test the global feature fusion of \eqref{eqn:hvf}, but only train the ResNet-3 with all available images in a \ac{US} study. In this way, it follows the spirit of Liu \etal{}'s global fusion strategy~\cite{liu2019ultrasound}. To reveal the impact of our hetero-fusion training strategy that uses different random combinations of \ac{US} images per epoch, we also test two \ac{HVF} variants, one using batch-normalization and one using instance-normalization. The latter helps measure the importance of using proper normalization strategies to manage the instability of \ac{HVF}'s batch statistics. Finally, we test our proposed model which incorporates \ac{VSP} on top of \ac{HVF} and clinical \ac{ROI} pooling. 

\textbf{Evaluation Protocols.} The problem setup is binary classification, \ie{} identifying patients with moderate to severe liver fibrosis, which are the patient cohorts requiring intervention. While we report full \acp{AUC}, we primarily focus on operating points within a useful range of specificity or precision. Thus, we  evaluate using partial \acp{AUC} that only consider false positive rates within $0$ to $30\%$  because higher values lose their practical usefulness. Partial \acp{AUC} are normalized to be within a range of $0$ to $1$. We also report recalls at a range of precision points (\acs{R@P90}, \acs{R@P85}, \acs{R@P80}) to reveal the achievable sensitivity at high precision points. We report mean values and mean graphs across all cross-validation folds.

\begin{figure}[t]
    \centering
    \includegraphics[width=.7\textwidth]{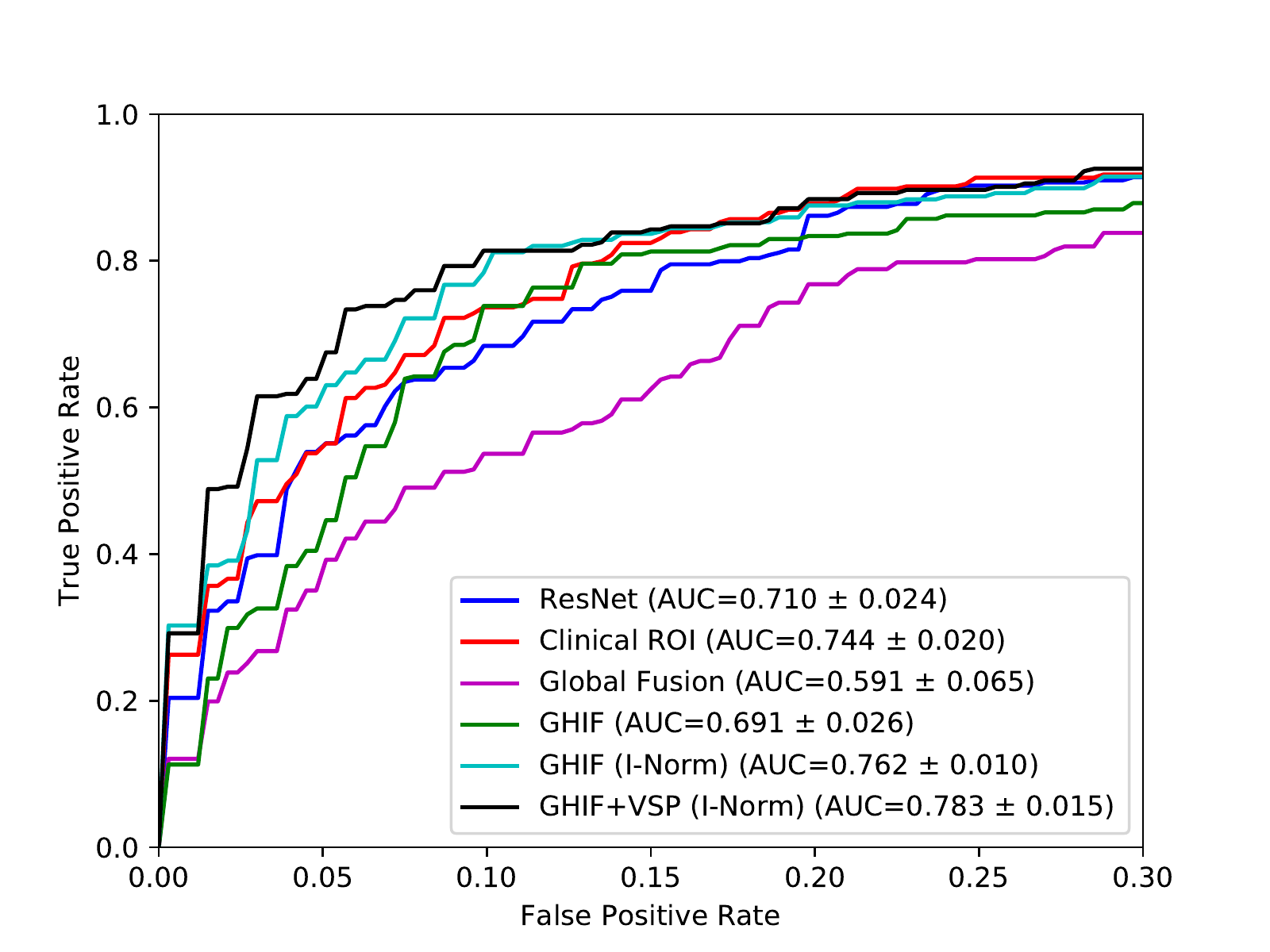}
    \caption{Partial \acs{ROC} curves are graphed, with corresponding partial \acsp{AUC} found in the legend. Partial \acs{AUC} scores have been normalized to range from $0-1$. Both \acsp{ROC} and \acsp{AUC} correspond to mean measures taken across the cross-validation folds.}
    \label{fig:aucs}
\end{figure}

\textbf{Results.} Tab.~\ref{tab1e-ablation} presents our \ac{AUC}, partial \ac{AUC} and recall values, whereas Fig.~\ref{fig:aucs} graphs the partial \acp{ROC}. Several conclusions can be drawn. First, clinical \ac{ROI} pooling produces significant boosts in performance, validating our strategy of forcing the network to focus on important regions of the image. 
Second, not surprisingly, global fusion without training with random combinations of images, performs very poorly, as only presenting all study images during training severely limits the data size and variability, handicapping the model. 
% Secondly, global fusion, which fuses features from all images in a study, but does not train with random combinations of said images, performs very poorly. This is not surprising, as only presenting all study images during training severely limits the data size and variability, handicapping the model. 
For instance, compared to variants that train on individual images, global fusion effectively reduces the training size by about a factor of $10$ in our dataset. In contrast, the \ac{HVF} variants, which train with the combinatorial number of random combinations of images, not only avoids drastically reducing the training set size, but can effectively increase it. Importantly, as the table demonstrates, using an appropriate choice of instance normalization is crucial in achieving good performance with \ac{HVF}. Although not shown, switching to instance normalization did not improve performance for the image-wise or global fusion models. The boosts in \ac{HVF} performance is apparent in the partial \ac{AUC} and recalls at high precision points, underscoring the need to analyze results at appropriate operating points. Finally, adding the \ac{VSP} provides even further performance improvements, particularly in \acs{R@P80}-\acs{R@P90} values, which see a roughly $4-7\%$ increase over \ac{HVF} alone. This indicates that \ac{VSP} can significantly enhance the recall at the very demanding precision points necessary for clinical use. In total, compared to a conventional classifier, the enhancements we articulate contribute to roughly $8\%$ improvements in partial \acp{AUC} and $22\%$ in \acs{R@P90} values. Table 1 of our supplementary material also presents \acp{AUC} when only choosing to input one particular view in the model during inference. We note that performance is highest when all views are inputted into the model, indicating that our pipeline is able to usefully exploit the information across views. Our supplementary also includes liver segmentation results and success and failure cases for our system. 

\section{Conclusion}
We presented a principled and effective pipeline for liver fibrosis characterization from \ac{US} studies, proposing several innovations: (1)  clinical \ac{ROI} pooling to discourage the network from focusing on spurious image features; (2) \ac{HVF} to manage any arbitrary number of images in the \ac{US} study in both training and inference; and (3) \ac{VSP} to tailor the analysis based on the liver view being presented using ``style''-based parameters. In particular, we are the first to propose a deep global hetero-fusion approach and the first to combine it with \ac{VSP}. Experiments demonstrate that our system can produce gains in partial \ac{AUC} and \acs{R@P90} of roughly $7\%$ and $22\%$, respectively on a dataset of $610$ patient studies. Future work should expand to other liver diseases and more explicitly incorporate other clinical markers, such as absolute or relative liver lobe sizing.

\bibliographystyle{splncs04}
\bibliography{refs}

\end{document}